\documentclass{ifacconf}

\usepackage{empheq}
\usepackage{changes}
\usepackage{cite}
\usepackage{graphicx}
\graphicspath{{Figs/}{../pdf/}{../jpeg/}}
\DeclareGraphicsExtensions{.pdf,.jpeg,.png}
\usepackage{lipsum}
\usepackage{algorithm}
\usepackage{array}
\usepackage{fixltx2e}
\usepackage{stfloats}
\usepackage{url}
\usepackage{booktabs}
\usepackage{enumerate}
\usepackage{setspace}

\usepackage{amsmath}
\usepackage{amssymb}
\usepackage{mathrsfs}

\definecolor{CBLUE}{RGB}{0,114,189}
\definecolor{CRED}{RGB}{217,83,25}
\definecolor{CYELLOW}{RGB}{237,177,32}
\definecolor{CPURPLE}{RGB}{126,47,142}

\usepackage{natbib}
\begin{document}
\begin{frontmatter}

\title{Quadratic Regularization of Data-Enabled Predictive Control: Theory and Application to Power Converter Experiments}


\author[ifa]{Linbin Huang,}
\author[ifa]{Jianzhe Zhen,}
\author[ifa]{John Lygeros,}
\author[ifa]{Florian D{\"o}rfler}
\address[ifa]{Automatic Control Laboratory, ETH, Z{\"u}rich 8092, Switzerland \text{(e-mail: \{linhuang, jizhen, jlygeros, dorfler\}@ethz.ch)}.}

\begin{abstract}
Data-driven control that circumvents the process of system identification by providing optimal control inputs directly from system data has attracted renewed attention in recent years. In this paper, we focus on understanding the effects of the regularization on the data-enabled predictive control (DeePC) algorithm. We provide theoretical motivation and interpretation for including a quadratic regularization term. Our analysis shows that the quadratic regularization term leads to robust and optimal solutions with regards to disturbances affecting the data. Moreover, when the input/output constraints are inactive, the quadratic regularization leads to a closed-form solution of the DeePC algorithm and thus enables fast calculations. On this basis, we propose a framework for data-driven synchronization and power regulations of power converters, which is tested by high-fidelity simulations and experiments.
\end{abstract}


\end{frontmatter}

\vspace{2.5mm}
\section{Introduction}
\vspace{-2.5mm}

Data-driven control is a promising technique for complex systems, as it does not rely on an accurate model of the system, which is in many cases difficult to obtain and maintain. For example, in power system applications, the system model is ever-changing due to different operation modes, uncertainties, and relaying. This poses great challenges for conventional model-based design and tuning, but may potentially be handled using data-driven control due to the great availability of data \citep{huang2019decentralized}.

The concept of data-driven control dates back to 1980s when the techniques of system identification and adaptive control became popular \citep{ljung1999system, goodwin2014adaptive}, and received renewed attention with the developments of iterative feedback tuning \citep{hjalmarsson1998iterative}, virtual reference feedback tuning \citep{campi2002virtual}, etc. Data-driven control based on reinforcement learning became a popular approach as well, which learns a policy directly from data \citep{recht2019tour}. It is also possible to provide robustness guarantees for data-driven control, e.g., by means of system-level synthesis \citep{boczar2018finite}.
In recent years, multiple data-driven control methods (e.g., \cite{coulson2019data, de2019formulas, berberich2020data}) have been proposed based on a result originally formulated by \cite{willems2005note} and recently extended by \cite{markovskyidentifiability} in the context of behavioral system theory. This result is also known as the {\em Fundamental Lemma} and shows that the subspace of input/output trajectories of a linear time-invariant (LTI) system can be obtained from the column span of a data Hankel matrix, thereby avoiding a parametric system representation.

Here we concentrate on a data-enabled predictive control (DeePC) algorithm proposed in \citep{coulson2019data}. Rather than a parametric model, in the spirit of the Fundamental Lemma, the DeePC algorithm relies only on input/output data measured from the unknown system to predict future trajectories and compute safe and optimal control inputs for the system. The DeePC algorithm has been successfully applied in many scenarios, including power systems \citep{huang2019data,huang2019decentralized}, motor drives \citep{carlet2020data}, and quadcopters \citep{elokda2019data}. It has been frequently observed that regularization terms are very important to ensure good performance when the system is subjected to disturbances. \cite{coulson2019regularized} showed that regularization provides distributional robustness against stochastic disturbances. \cite{xue2020data} arrive at similar conclusions via a robust control formulation. Though, the robust performance still needs to be investigated thoroughly under bounded disturbances.

In this paper, we provide theoretical motivation and interpretation for adding a quadratic regularization term in DeePC from a robust optimization perspective. We show how including a quadratic regularization is equivalent to a min-max optimization problem that minimizes the worst-case performance for bounded disturbance sets imposed on the input/output data. This suggests that, unlike system identification methods that filter out the disturbances to fit a certain model, the regularized DeePC algorithm {directly provides robust control inputs to the system with regards to the disturbances in data.} Moreover, the quadratic regularization allows one to obtain a closed-form solution of the DeePC algorithm enabling the fast calculation of optimal inputs.
We exploit these properties of DeePC with quadratic regularization to robustly control a grid-connected power converter to achieve data-driven synchronization and power regulation. Our results are illustrated with high-fidelity simulations and experiments.

\section{Data-Enabled Predictive Control}

\subsection{Notation and Preliminaries on the Fundamental Lemma}

Consider an $n{\rm th}$-order discrete-time LTI system
\begin{equation}
\left\{ \begin{array}{l}
{x_{t + 1}} = A{x_t} + B{u_t}\\
{y_t} = C{x_t} + D{u_t},
\end{array} \right.\,		\label{eq:ABCD}
\end{equation}
where $A \in \mathbb{R}^{n \times n}$, $B \in \mathbb{R}^{n \times m}$, $C \in \mathbb{R}^{p \times n}$, $D \in \mathbb{R}^{p \times m}$, $x_t \in \mathbb{R}^n$ is the state of the system at $t \in \mathbb{Z}_{ \ge 0}$, $u_{t} \in \mathbb{R}^m$ is the input vector, and $y_{t} \in \mathbb{R}^p$ is the output vector.

The \textit{lag} of the system (\ref{eq:ABCD}) is defined by the smallest integer $\ell \in \mathbb{Z}_{ \ge 0}$ such that the observability matrix
\vspace{-0.5mm}
\begin{equation*}
\mathscr{O}_{\ell}(A,C) := {\rm{col}}(C,CA,...,CA^{\ell-1})
\vspace{-0.5mm}
\end{equation*}
has rank $n$, i.e., the state can be reconstructed from $\ell$ measurements. Here ${\rm{col}}(a_0,a_1,...,a_i):=[a_0^{\top}\; a_1^{\top}\; \cdots \;a_i^{\top}]^{\top}$.
In a data-driven setting, $\ell$ and $n$ are generally not known, but upper bounds can usually be inferred. 


Consider $L,T \in \mathbb{Z}_{ \ge 0}$ with $T \geq L > \ell$ and length-$T$ input and output trajectories of \eqref{eq:ABCD}: $u = {\rm{col}}(u_0,u_1,\dots u_{T-1})\in \mathbb{R}^{mT}$ and $y = {\rm{col}}(y_0,y_1, \dots y_{T-1})\in \mathbb{R}^{pT}$. We denote the set of all such trajectories as the {\em restricted behavior}. For the inputs $u$, define the Hankel matrix of depth $L$ as
\begin{equation}
\mathscr{H}_L(u) := \left[ {\begin{array}{*{20}{c}}
	{{u_0}}&{{u_1}}& \cdots &{{u_{T - L}}}\\
	{{u_1}}&{{u_2}}& \cdots &{{u_{T - L + 1}}}\\
	\vdots & \vdots & \ddots & \vdots \\
	{{u_{L-1}}}&{{u_{L}}}& \cdots &{{u_{T-1}}}
	\end{array}} \right] \,.		
\label{eq:Hankel_L}
\end{equation}
Accordingly, for the outputs define the Hankel matrix $\mathscr{H}_L(y)$. Consider the stacked matrix
$\mathscr{H}_L(u,y) = \left[\begin{smallmatrix}\mathscr{H}_L(u)\\\mathscr{H}_L(y)\end{smallmatrix}\right]$.

By \cite[Corollary 19]{markovskyidentifiability}, the length-$L$ restricted behavior equals the image of $\mathscr{H}_L(u,y)$ if and only if rank$\left(\mathscr{H}_L(u,y)\right)=mL + n$. In words, the Hankel matrix $\mathscr{H}_L(u,y)$ composed of a single length-$T$  trajectory parametrizes all length-$L$ trajectories provided that (and only that) rank$\left(\mathscr{H}_L(u,y)\right)=mL + n$.
This result extends and includes the original {\em Fundamental Lemma}  \cite[Theorem 1]{willems2005note} which requires controllability and persistency of excitation of order $L+n$ (i.e., $\mathscr{H}_{L+n}(u)$ must have full row rank) as sufficient conditions. The result by \cite{markovskyidentifiability} also extends to mosaic Hankel, Page, and trajectory matrix structures.

These behavioral results can be leveraged for data-driven prediction and estimation as follows. Consider $T_{\rm ini},N,T \in \mathbb{Z}_{ \ge 0}$ as well as an input/output time series ${\rm{col}}(u^{\rm{d}},y^{\rm{d}}) \in \mathbb{R}^{(m+p)T}$ so that rank$\left(\mathscr{H}_{T_{\rm ini}+N}(u^{\rm{d}},y^{\rm{d}})\right)=m(T_{\rm ini}+N) + n$. Here the superscript ``d'' denotes data collected offline {in a single shot}, and the rank condition can be met by choosing $u^{\rm{d}}$ to be  persistently exciting of sufficiently high order {(assuming that the system is controllable)}. Further partition the Hankel matrix $\mathscr{H}_{T_{\rm ini}+N}(u^{\rm{d}},y^{\rm{d}})$ as
\begin{equation}
\left[ {\begin{array}{*{20}{c}}
	{{U_{\rm P}}}\\
	{{U_{\rm F}}}
	\end{array}} \right] := \mathscr{H}_{T_{\rm ini}+N}(u^{\rm{d}})\,,\;\;\left[ {\begin{array}{*{20}{c}}
	{{Y_{\rm P}}}\\
	{{Y_{\rm F}}}
	\end{array}} \right] := \mathscr{H}_{T_{\rm ini}+N}(y^{\rm{d}})\,,		\label{eq:partition_Huy}
\end{equation}
where $U_{\rm P} \in \mathbb{R}^{mT_{\rm ini} \times H}$, $U_{\rm F} \in \mathbb{R}^{mN \times H}$, $Y_{\rm P} \in \mathbb{R}^{pT_{\rm ini} \times H}$, $Y_{\rm F} \in \mathbb{R}^{pN \times H}$, and $H = T-T_{\rm ini}-N+1$.  In the sequel, the data in the partition with subscript P (for ``past'') will be used to estimate the initial condition of the system, whereas the data with subscript F will be used to predict the ``future'' trajectories. In this case, $T_{\rm ini}$ is the length of an initial trajectory measured in the immediate past during on-line operation, and $N$ is the length of a predicted trajectory starting from the initial trajectory. Recall that the image of $\mathscr{H}_{T_{\rm ini}+N}(u^{\rm{d}},y^{\rm{d}})$ spans all length-$(T_{\rm ini}+N)$ trajectories, that is,
 ${\rm{col}}(u_{\rm ini},u,y_{\rm ini},y) \in \mathbb R^{(m+p)(T_{\rm ini}+N)}$ is a trajectory of (\ref{eq:ABCD}) if and only if there exists $g \in \mathbb{R}^{H}$ so\,that
\begin{equation}
\left[ {\begin{array}{*{20}{c}}
	{{U_{\rm P}}}\\
	{{Y_{\rm P}}}\\
	{{U_{\rm F}}}\\
	{{Y_{\rm F}}}
	\end{array}} \right]g = \left[ {\begin{array}{*{20}{c}}
	{{u_{\rm ini}}}\\
	{{y_{\rm ini}}}\\
	u\\
	y
	\end{array}} \right]\,.		\label{eq:Hankel_g}
\end{equation}
The initial trajectory ${\rm{col}}(u_{\rm ini},y_{\rm ini}) \in \mathbb R^{(m+p)T_{\rm ini}}$ can be thought of as setting the initial condition for the future (to be predicted) trajectory ${\rm{col}}(u,y)\in \mathbb R^{(m+p)N}$. In particular, if $T_{\rm ini} \ge \ell$, for every given future input trajectory $u$, the future output trajectory $y$ is uniquely determined through (\ref{eq:Hankel_g}) \citep{markovsky2008data}.

\subsection{Review of the DeePC algorithm}

The DeePC algorithm proposed in \cite{coulson2019data} directly uses input/output data collected from the unknown system to predict the future behaviour, and perform optimal and safe control without identifying a parametric system representation. More specifically, DeePC solves the following optimization problem to obtain the optimal future control inputs
\begin{equation}
\begin{array}{cl}
\mathop {{\rm{min}}}\limits_{g,\sigma_y \atop u \in \mathcal U, y \in \mathcal Y} & \;\;{\left\| u \right\|_R^2} + {\left\| {y - r} \right\|_Q^2} + {\lambda _y}{\left\| \sigma_y \right\|_2^2} \\
{\rm s.t.} & \;\;\left[ {\begin{array}{*{20}{c}}
	{{U_{\rm P}}}\\
	{{Y_{\rm P}}}\\
	{{U_{\rm F}}}\\
	{{Y_{\rm F}}}
	\end{array}} \right]g = \left[ {\begin{array}{*{20}{c}}
	{{u_{\rm ini}}}\\
	{{y_{\rm ini}}}\\
	u\\
	y
	\end{array}} \right] + \left[ {\begin{array}{*{20}{c}}
	0\\
	\sigma_y\\
	0\\
	0
	\end{array}} \right]\,,		
\label{eq:DeePC}
\end{array}
\end{equation}
where the sets $\mathcal U \subseteq \mathbb{R}^{mN}$ and $\mathcal Y \subseteq \mathbb{R}^{pN}$ describe the feasible region of the input $u$ and output $y$ of the system (assumed to be polytopes), $R \in \mathbb{R}^{mN \times mN}$ is the control cost matrix (positive definite), $Q \in \mathbb{R}^{pN \times pN}$ is the output cost matrix (positive semidefinite), $\sigma_y \in \mathbb{R}^{pT_{\rm ini}}$ is an auxiliary slack variable to ensure feasibility of the least-square initial condition estimation, $\lambda_y \in \mathbb{R}_{ > 0}$ is regularization parameter, $r \in \mathbb{R}^{pN}$ is the reference trajectory for the outputs, $N$ is the prediction horizon, ${\rm{col}}(u_{\rm ini},y_{\rm ini})$ consists of the most recent input/output trajectory of (\ref{eq:ABCD}) of length $T_{\rm ini}$, and ${\left\| a \right\|_X^2}$ denotes the quadratic form $a^\top Xa$.

DeePC involves solving the convex optimization problem (\ref{eq:DeePC}) in a receding horizon manner, that is, after calculating the optimal control sequence $u^\star$, we apply $(u_t,...,u_{t+k-1}) = (u_0^{\star},...,u_{k-1}^{\star})$ to the system for $k \le N-1$ time steps, then, reinitialize the problem (\ref{eq:DeePC}) by updating ${\rm{col}}(u_{\rm ini},y_{\rm ini})$ to the most recent input and output measurements, and setting $t$ to $t+k$, to calculate the new optimal control for the next $k \le N-1$ time steps. As usual in MPC, the control horizon $k$ is a design parameter.

A regularization term on $g$ can be included in the cost function of \eqref{eq:DeePC} to ensure robustness in case of noisy data samples. For example, when stochastic disturbances affect the output measurements, a two-norm regularization on $g$ coincides with distributional two-norm robustness in the trajectory space \citep{coulson2019regularized}. Other forms of regularization (one-norm penalty, etc.) on $g$ can also be used to provide different types of robustness. It was also reported in \citep{xue2020data} that a robust control formulation of \eqref{eq:DeePC} leads to a two-norm regularization.

\section{Robustness Induced by Regularization}

In this section, we provide a theoretical motivation and interpretation for adding a quadratic regularization on $g$ in DeePC from a robust optimization perspective.
We start by noting that the explicit equality constraints in~\eqref{eq:DeePC} can be eliminated, leading to
\begin{equation}
\begin{array}{l}
\mathop {{\rm{min}}}\limits_{g \in \mathcal{G}} \;\;{\left\| U_{\rm F}g \right\|_R^2} + {\left\| {Y_{\rm F}g - r} \right\|_Q^2} + \lambda_y\left\|Y_{\rm P}g-y_{\rm ini}\right\|_2^2 \,,
\label{eq:DeePC1}
\end{array}
\end{equation}
where $\mathcal{G} = \{g \in \mathbb{R}^{H} \ | \ U_{\rm F}g \in \mathcal U, \ Y_{\rm F}g \in \mathcal Y, \ U_{\rm P}g = u_{\rm ini}\}$ {is assumed to be nonempty}. We rewrite \eqref{eq:DeePC1} as
\begin{equation}
\begin{array}{l}
\mathop {{\rm{min}}}\limits_{g \in \mathcal{G}} \;\;\left\|Ag-b\right\|_2^2 \,,
\label{eq:DeePC2}
\end{array}
\end{equation}
where $\mathcal{G} = \{g \ | \ G g \le q \}$ for some $G$ and $q$,
\begin{equation*}
A = \left[ {\begin{array}{*{20}{c}}
	\lambda_y^{\frac{1}{2}}  Y_{\rm P}\\
	R^{\frac{1}{2}}  U_{\rm F}\\
	Q^{\frac{1}{2}}  Y_{\rm F}
	\end{array}} \right] \quad \text{and} \quad b = \left[ {\begin{array}{*{20}{c}}
	\lambda_y^{\frac{1}{2}}  y_{\rm ini}\\
	0\\
	Q^{\frac{1}{2}}r
	\end{array}} \right] \,.
\end{equation*}

Notice that the solution of \eqref{eq:DeePC2} will in general not be robust to disturbances on $A$, which in our case contains the historical input/output data. As a remedy, one can instead consider a robust version of \eqref{eq:DeePC2} by solving the following min-max optimization problem
\begin{equation}\label{eq:DeePC_robust}
\mathop {{\rm{min}}}\limits_{g \in \mathcal{G}} \;\;\mathop {{\rm{max}}}\limits_{\|\Delta\|_F \le \beta}\; \left\|(A + \Delta)g- b \right\|_2\,
\end{equation}
where $\Delta$ is the disturbance matrix affecting $A$, $\|\Delta\|_F = \sqrt{ \sum_{i,j} \Delta_{ij}^2 }$ denotes the Frobenius norm, and $\beta > 0$ is the bound on $\|\Delta\|_F$.
Eq.~\eqref{eq:DeePC_robust} minimizes the worst-case cost with respect to the disturbances on the matrix $A$.
Hence, the minimizer $g^\star$ (and thus the optimal control input sequence $u^\star = U_{\rm F}g^\star$) is robust against disturbances affecting the data matrices ($U_{\rm P}$, $Y_{\rm P}$, and $Y_{\rm F}$).

The following result explicitly shows how the minimizer of \eqref{eq:DeePC_robust} can be obtained by solving a regularized version of \eqref{eq:DeePC2} with a quadratic regularization on $g$.

\vspace{3mm}

\begin{thm}
\label{thm1}
(Robustness from quadratic regularization). If~$g^\star \in \mathbb{R}^{H}$ is a minimizer of
\begin{equation}\label{eq:DeePC_quadratic}
\mathop {{\rm{min}}}\limits_{g \in \mathcal{G}} \; \left\|Ag- b \right\|_2^2 + \lambda_g \left\|g \right\|_2^2, \,
\end{equation}
then $g^\star$ minimizes \eqref{eq:DeePC_robust} with
\begin{equation}\label{eq:lambda_g}
	\beta =  \begin{cases}
		\frac{\lambda_g \left\| g^\star\right\|_2}{\left\|Ag^\star - b\right\|_2} & \text{if $Ag^\star \ne b$} \\
		\lambda_g \left\| g^\star\right\|_2 & \text{otherwise.}
	\end{cases}
\end{equation}
Moreover, if $g^\star \neq 0$, $\beta$ in \eqref{eq:lambda_g} is strictly monotonic increasing with $\lambda_g$ chosen in \eqref{eq:DeePC_quadratic}.
\end{thm}
\begin{pf}
Consider the Lagrangian of \eqref{eq:DeePC_quadratic}
\begin{equation}\label{eq:L_Q}
	\mathcal{L_Q}(g,\mu_Q) = \left\|Ag-b\right\|_2^2 + {\lambda_g}{\left\| g \right\|_2^2} + \mu_Q^\top(Gg - q)\,,
\end{equation}
where $\mu_Q$ is the vector of the dual variables.  Since~$\mathcal{G}$ is nonempty, there exists a solution $(g^\star,\mu_Q^\star)$ to the Karush–Kuhn–Tucker (KKT) conditions of \eqref{eq:DeePC_quadratic}
\begin{empheq}[left={\empheqlbrace}]{align}
	& 2{A^\top (Ag-b)} + 2{\lambda_g g} +   G^\top \mu_Q = 0    \,,          \label{eq:KKT_LQ1}\\
	& \mu_Q^\top (Gg - q) = 0  \,,   \label{eq:KKT_LQ2}  \\
	& Gg \le  q            \,,     \label{eq:KKT_LQ3}  \\
	& \mu_Q \ge 0          \,.       \label{eq:KKT_LQ4}
\end{empheq}
Therefore, the vector $g^\star$ is a minimizer of \eqref{eq:DeePC_quadratic}.

Following \cite[Theorem~2]{bertsimas2018characterization} and \citep{el1997robust}, the min-max problem~\eqref{eq:DeePC_robust} can be equivalently reformulated as
\begin{equation}\label{eq:robust_min}
 \mathop {{\rm{min}}}\limits_{g \in \mathcal{G}} \;\;\left\|Ag-b\right\|_2 + {\beta}{\left\| g \right\|_2} \,.
\end{equation}
Consider the Lagrangian of \eqref{eq:robust_min}
\begin{equation}\label{eq:L}
\mathcal{L}(g,\mu) = \left\|Ag-b\right\|_2 + {\beta}{\left\| g \right\|_2} + \mu^\top(Gg - q)\,,
\end{equation}
where $\mu$ is the vector of the dual variables. By choosing $\beta$ in \eqref{eq:lambda_g}, it can be verified that $(g^\star,  \mu^\star, y^\star, z^\star)$, where
\begin{equation*}
	(\mu^\star,  y^\star)  = \begin{cases} \left( \frac{\mu_Q^\star}{2 \| Ag^\star - b \|_2}, \frac{A^\top (A g^\star - b)}{\| A g^\star - b \|_2} \right) & \text{if $Ag^\star \ne b$}\\ \left(\frac{\mu_Q^\star}{2} , 0 \right) & \text{otherwise,} \end{cases}
\end{equation*}
$z^\star =  \beta g^\star / \| g^\star \|_2$ if $g^\star \ne 0$, and $z^\star = 0$ otherwise, and $g^\star$ as before,
satisfy the KKT conditions of \eqref{eq:robust_min}:
\begin{empheq}[left={\empheqlbrace}]{align}
& y + z +   G^\top \mu = 0 \,, \label{eq:KKT_L1}\\
& \|A \bar g-b\|_2 \ge \|A g-b\|_2  + y^\top (\bar g - g), & \forall \bar g \in \mathbb R^H  \label{eq:KKT_L1a} \\
& \beta \|\bar g\|_2 \ge \beta \|g\|_2 + z^\top (\bar g - g),  & \forall \bar g \in \mathbb R^H  \label{eq:KKT_L1b} \\
& \mu^\top (Gg - q) = 0  \,,   \label{eq:KKT_L2}  \\
& Gg \le  q            \,,     \label{eq:KKT_L3}  \\
& \mu \ge 0            \,,     \label{eq:KKT_L4}
\end{empheq}
where \eqref{eq:KKT_L1a} and \eqref{eq:KKT_L1b} follow from the definition of subgradient for the convex functions $\|Ag-b\|_2$ and $\|g\|_2$, respectively.
Thus, the vector $g^\star$ is also a minimizer of \eqref{eq:robust_min} (and \eqref{eq:DeePC_robust}).

Next we prove the monotonic relationship between $\beta$ and $\lambda_g$. Let $g_1$ be the minimizer of \eqref{eq:DeePC_quadratic} with $\lambda_g = \lambda_{g1} >0$ (and the minimizer of \eqref{eq:robust_min} with $\beta = \beta_1$), and $g_2$ the minimizer of \eqref{eq:DeePC_quadratic} with $\lambda_g = \lambda_{g2} > \lambda_{g1}$ (and the minimizer of \eqref{eq:robust_min} with $\beta = \beta_2$). If $g_1 \neq g_2$, according to the definitions of $g_1$ and $g_2$, we have
\begin{equation}\label{eq:g_a}
\left\|Ag_1-b\right\|_2^2 + \lambda_{g1}{\left\| g_1 \right\|_2^2} - \lambda_{g1}{\left\| g_2 \right\|_2^2} < \left\|Ag_2-b\right\|_2^2 \,,
\end{equation}
\begin{equation}\label{eq:g_b}
\left\|Ag_2-b\right\|_2^2 < \left\|Ag_1-b\right\|_2^2 + \lambda_{g2}{\left\| g_1 \right\|_2^2} - \lambda_{g2}{\left\| g_2 \right\|_2^2} \,,
\end{equation}
leading to
\begin{equation}\label{eq:g_ab}
\lambda_{g1}({\left\| g_1 \right\|_2^2} - {\left\| g_2 \right\|_2^2}) < \lambda_{g2}({\left\| g_1 \right\|_2^2} - {\left\| g_2 \right\|_2^2}) \,,
\end{equation}
which indicates that ${\left\| g_1 \right\|_2^2} > {\left\| g_2 \right\|_2^2}$ because $ 0 < \lambda_{g1} < \lambda_{g2}$. Then, we have ${\left\| g_1 \right\|_2} > {\left\| g_2 \right\|_2}$. According to the definitions of $\beta_1$ and $\beta_2$, we also have
\begin{equation}\label{eq:beta_a}
\left\|Ag_1-b\right\|_2 + \beta_1{\left\| g_1 \right\|_2} - \beta_1{\left\| g_2 \right\|_2} < \left\|Ag_2-b\right\|_2 \,,
\end{equation}
\begin{equation}\label{eq:beta_b}
\left\|Ag_2-b\right\|_2 < \left\|Ag_1-b\right\|_2 + \beta_2{\left\| g_1 \right\|_2} - \beta_2{\left\| g_2 \right\|_2} \,.
\end{equation}
It can then be deduced that $\beta_1 < \beta_2$. If $g_1 = g_2 \neq 0$, we have $\beta_1 < \beta_2$ according to \eqref{eq:lambda_g}. Hence, $\beta$ is increasing with the increase of $\lambda_g$ if $g^\star \neq 0$. This completes the proof.
\end{pf}\vspace{-0.8mm}

Theorem~\ref{thm1} shows that the robust solution of \eqref{eq:DeePC2} can be obtained by simply adding a quadratic regularization term, and thus the problem can be solved efficiently.
Moreover, the Frobenius norm of the disturbance matrix is bounded by $\beta$, which is monotonically increasing in $\lambda_g$.
The bound $\beta$ is an a-posteriori guarantee obtained after solving \eqref{eq:DeePC_quadratic}.
With a larger $\lambda_g$, the disturbance set becomes larger including more possible disturbances. Obviously, an overly large $\lambda_g$ gives a conservative formulation.

In Theorem~\ref{thm1}, we consider disturbances affecting $A$ (i.e., affecting the data matrices $U_{\rm P}$, $Y_{\rm P}$, and $Y_{\rm F}$). The following result will show that the quadratic regularization on $g$ also provides robustness against disturbances on the vector $b$.


\begin{cor}
\label{cor1}
If a vector $g^\star \in \mathbb{R}^{H}$ is a minimizer of \eqref{eq:DeePC_quadratic}, then $g^\star$ minimizes \begin{equation}\label{eq:DeePC_minimax1}
\mathop {{\rm{min}}}\limits_{g \in \mathcal{G}} \;\;\mathop {{\rm{max}}}\limits_{\left\|[\Delta\;\xi]\right\|_F \le \beta'}\; \left\|(A + \Delta)g- (b+\xi) \right\|_2\,
\end{equation}
with
\begin{equation}\label{eq:lambda_g1}
\beta' =  \begin{cases}
		\frac{\lambda_g \sqrt{\left\| g^\star\right\|_2^2+1}}{\left\|Ag^\star - b\right\|_2} & \text{if $Ag^\star \ne b$} \\
		\lambda_g \sqrt{\left\| g^\star\right\|_2^2+1} & \text{otherwise,}
	\end{cases}
\end{equation}
where $\Delta$ is the disturbance matrix affecting $A$, and $\xi$ is the disturbance vector affecting $b$.
\end{cor}

\begin{pf}
Consider an augmented decision variable $\bar g  = {\rm col}(g,-1)$. If $g^\star \in \mathbb{R}^{H}$ is a minimizer of \eqref{eq:DeePC_quadratic}, then $\bar g^\star = {\rm col}(g^\star,-1)$ minimizes
\begin{equation}\label{eq:DeePC3}
\mathop {{\rm{min}}}\limits_{\bar g  \in \mathcal{G}'} \;\;\left\|A'\bar g \right\|_2^2 + {\lambda_g}{\left\| \bar g \right\|_2^2} \,,
\end{equation}
where $\mathcal{G}' = \{\bar g \ | \ \bar g = {\rm col}(g,-1), \ g \in \mathcal{G}\}$, and $A' = [A\;b]$. Then, it follows from Theorem~\ref{thm1} that $\bar g^\star$ minimizes
\begin{equation}\label{eq:DeePC_minimax2}
\mathop {{\rm{min}}}\limits_{ \bar g \in \mathcal{G}'} \;\;\mathop {{\rm{max}}}\limits_{\left\|\Delta'\right\|_F \le \beta'}\; \left\|(A' + \Delta')\bar g \right\|_2\,,
\end{equation}
{with $\beta'$ given in~\eqref{eq:lambda_g1}.}
Then, \eqref{eq:DeePC_minimax2} can be rewritten as \eqref{eq:DeePC_minimax1} with $\Delta' = [\Delta\;\xi]$, that is, $g^\star$ minimizes \eqref{eq:DeePC_minimax1}. 
\end{pf}\vspace{1.5mm}

Corollary~\ref{cor1} indicates that the quadratic regularization in \eqref{eq:DeePC_quadratic} actually provides robustness against disturbances affecting both $A$ and $b$. In other words, the solution to \eqref{eq:DeePC_quadratic} is robust to the disturbances affecting the data matrices ($U_{\rm P}$, $Y_{\rm P}$, and $Y_{\rm F}$) and the initial output trajectory $y_{\rm ini}$. However, the considered disturbance set (resulting from the quadratic regularization) is conservative because some entries in $A$ (e.g., the entries of $U_{\rm P}$) and $b$ (e.g., the entries of $r$) are not subjected to disturbances. Moreover, how the considered disturbances affect input/output signals is also related to the scaling matrices, e.g., $Q^\frac{1}{2}$ in $A$. We will consider structured disturbance sets in future work.

\section{Applications to Power converters}

Power converters have been extensively used in modern power systems to accommodate large-scale renewable generation and high-voltage DC systems.
Conventional control schemes for power converters are model-based, however, an accurate system model can rarely be obtained especially when the converter is connected to a power grid. This may result in inferior performance or even instabilities. In this section, we apply the DeePC algorithm with quadratic regularization to perform model-free optimal control in grid-connected power converters.

The DeePC algorithm has been successfully applied in grid-connected power converters for data-driven oscillation damping in \cite{huang2019data, huang2019decentralized}. However, in these works, the converters still rely on conventional schemes to achieve grid synchronization, power regulation, etc., and the DeePC algorithm only provides the control signal to stabilize the system.
Here, we present a data-driven control framework for power converters to achieve grid synchronization and power regulation simultaneously.

\subsection{Data-Driven Synchronization and Power Regulation}

\begin{figure}
\begin{center}
\includegraphics[width=8.4cm]{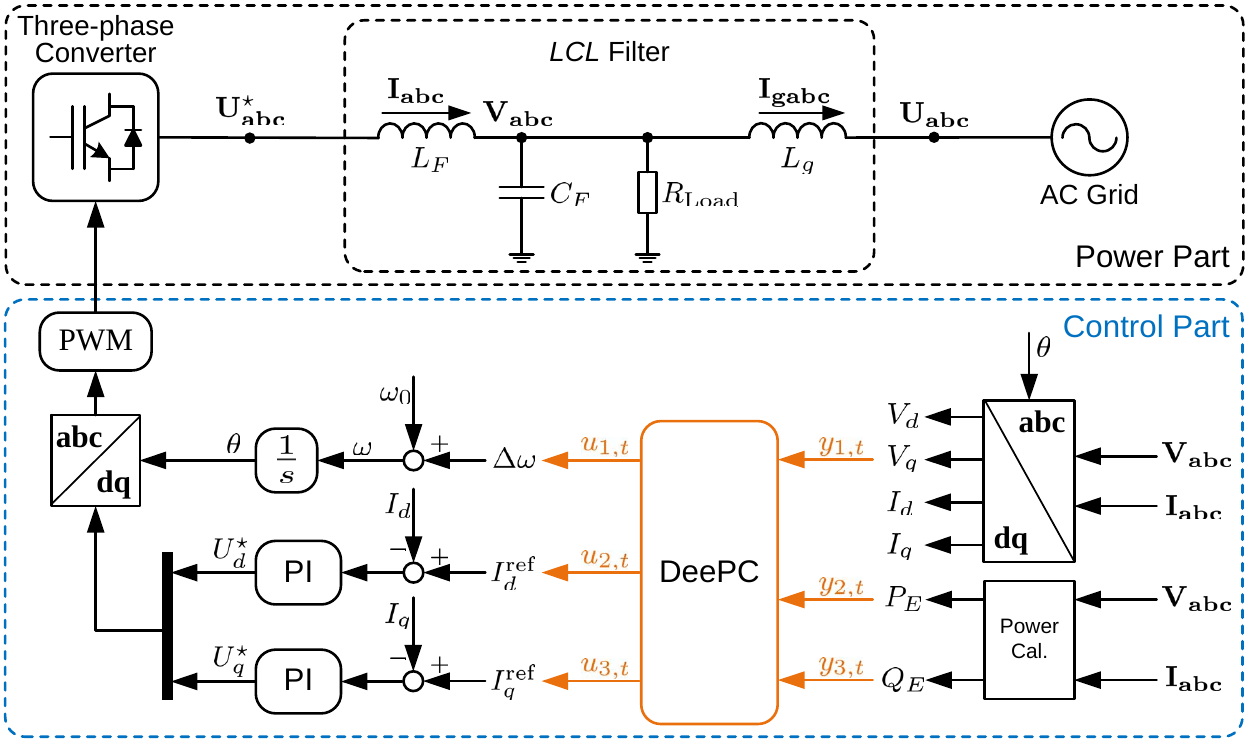}
\vspace{-2mm}
\caption{A data-driven control framework for power converters based on DeePC algorithm.}
\vspace{2mm}
\label{Fig_control_scheme}
\end{center}
\end{figure}

Fig.~\ref{Fig_control_scheme} shows a three-phase power converter which is connected to an AC power grid via an {\em LCL} filter. A current control loop is used to enable fast current tracking and current limitation. The control objectives of the converter include grid synchronization and active/reactive power regulation.
As illustrated in Fig.~\ref{Fig_control_scheme}, we choose the {\em q}-axis voltage component $V_q$, active power $P_E$, and reactive power $Q_E$ to be the output signals of the converter system. The DeePC algorithm provides optimal control inputs that enter the converter system as internal frequency deviation $\Delta \omega$ and current references $I_d^{\rm ref}$, $I_q^{\rm ref}$. Note that $u_{i,t}$ is the $i{\rm th}$ element of $u_t$ and $y_{i,t}$ the $i{\rm th}$ element of $y_t$ in Fig.~\ref{Fig_control_scheme}.

Before the DeePC algorithm is activated, input/output data measured from the converter system needs to be collected to construct the Hankel matrices.
Note that a phase-locked loop (PLL) is used to stabilize the system when the DeePC is not activated, but is then bypassed once the DeePC is activated.
After constructing the Hankel matrices, the optimal control inputs are obtained by solving \eqref{eq:DeePC_quadratic} in a receding-horizon manner. The control framework in Fig.~\ref{Fig_control_scheme} achieves grid synchronization by minimizing the trajectory of $V_q$ (i.e., making the control coordinate aligned with the voltage vector). The active/reactive power regulations are achieved by giving references for the output signals $P_E$ and $Q_E$. We set the reference vector to $r = I_N \otimes {\rm col}(0,P_0,Q_0)$ to achieve the above control objectives, where $\otimes$ denotes the Kronecker product, $I_N \in \mathbb{R}^{N \times N}$ is the identity matrix, $P_0$ is the active power reference value, and $Q_0$ is the reactive power reference value.

\subsection{Closed-Form Solutions}

The quadratic regularization in \eqref{eq:DeePC_quadratic} not only provides robustness for the DeePC algorithm, but also allows one to obtain the solution by quadratic programming. Moreover, if there are no input or output constraints (or if they are inactive at the optimal solution \eqref{eq:closed_form} below), we can compactly obtain the closed-form solution of \eqref{eq:DeePC_quadratic} as
\begin{equation}\label{eq:closed_form}
g^\star = M\left[ {\begin{array}{*{20}{c}}
	2A^\top b\\
	u_{\rm ini}
	\end{array}} \right]
\end{equation}
where $M \in \mathbb{R}^{H \times (H+mT_{\rm ini})}$ is defined by
\begin{equation*}
\left[ {\begin{array}{*{20}{c}}
	M \\
	\bar M
	\end{array}} \right] = \left[ {\begin{array}{*{20}{c}}
	2 (A^\top A + \lambda_g I_H)&U_{\rm P}^\top\\
	U_{\rm P} &\bf{0}
	\end{array}} \right]^{-1}\,.
\end{equation*}

The closed-form solution in \eqref{eq:closed_form} is simply a linear mapping of the initial trajectory ${\rm col}(u_{\rm ini}, y_{\rm ini})$ and the reference vector $r$~\citep{carlet2020data,alexandru2020data}. If it happens to satisfy any input and output constraints, it provides a fast way to obtain the minimizer of \eqref{eq:DeePC_quadratic} and thus the optimal input sequence $u^\star  = U_{\rm F}g^\star$.

\subsection{Simulation Results}

In our simulations, we use the base values $f_{\rm b} = 50{\rm Hz}$, $S_{\rm b} = 1.50{\rm kW}$, and $U_{\rm b} = 208{\rm V}$ for per-unit calculations. The {\em LCL} parameters are: $L_F = 0.04{\rm p.u.}$ (with resistance $R_F = 0.03{\rm p.u.}$), $L_g = 0.04{\rm p.u.}$ (with resistance $R_g = 0.03{\rm p.u.}$), and capacitor $C_F = 0.02{\rm p.u.}$. The local load is $R_{\rm Load} = 2.23{\rm p.u.}$. The PI parameters of the current control are $\{0.15,10\}$.
The parameters for DeePC are: $T_{\rm ini} = 6$, $N = 12$, $T = 500$, $\lambda_y = 10^4$, $R = I_{3N}$, and $Q = 400I_{3N}$. The input/output constraints are ignored. The control horizon is $k=1$, that is, after applying $u^\star_0$ to the system, we update the initial trajectory and then calculate the optimal inputs for the next time step.

\begin{figure}
\begin{center}
\includegraphics[width=8.4cm]{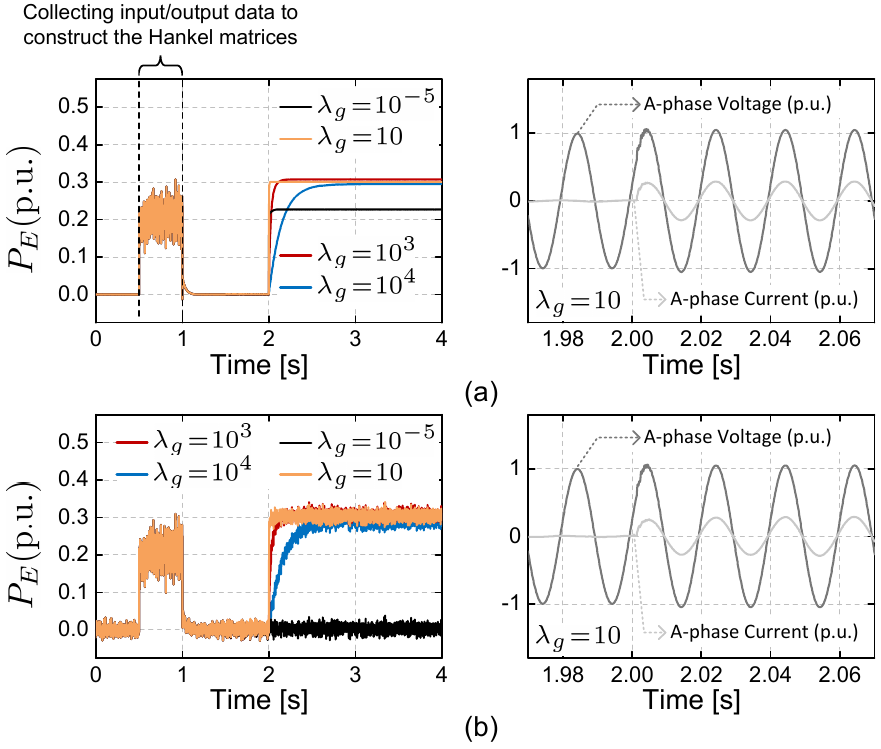}
\vspace{-3mm}
\caption{Active power responses of the converter with different values of $\lambda_g$. (a) Without measurement noise. (b) With measurement noise power being $10^{-7}$ (p.u.).}
\vspace{2mm}
\label{Fig_sim}
\end{center}
\end{figure}

Fig.~\ref{Fig_sim} shows the time-domain active power responses of the converter. Before $0.5 {\rm s}$, the system operates with $I_d^{\rm ref} = I_q^{\rm ref} = 0$, and a PLL is used to generate $\Delta \omega$.
From $0.5 {\rm s}$ to $1{\rm s}$, the input/output data of the converter is collected to construct the Hankel matrices. During this period, the output of the PLL and a white-noise signal are added to the input signal $\Delta \omega$, i.e., $u_{1,t}$; a constant setting value $0.2$ and a white-noise signal are added to $I_d^{\rm ref}$, i.e., $u_{2,t}$; a constant setting value $0$ and a white-noise signal are added to $I_q^{\rm ref}$, i.e., $u_{3,t}$. The white-noise signals (with noise power $6 \times 10^{-7} {\rm p.u.}$) ensure that the input trajectory is persistently exciting. After collecting the data, the noise is switched off, and $I_d^{\rm ref}$ is set back to 0. The PLL is then switched off and the DeePC controller is switched on at $1.5 {\rm s}$ with $Q_0 = 0$ and $P_0 = 0$. Finally, active power reference $P_0$ steps from $0$ to $0.3{\rm p.u.}$ at $2{\rm s}$.

In Fig.~\ref{Fig_sim}~(a), the measurement noise for the output signals is not considered, and the converter has very fast and smooth response to the change of $P_0$ (without any overshoot) when choosing $\lambda_g = 10$. Moreover, the active power perfectly tracks the reference value in steady state, which is achieved by penalizing the tracking error in the cost function of \eqref{eq:DeePC_quadratic}. The response becomes slower by increasing $\lambda_g$ to $10^3$, and it becomes even slower by further increasing $\lambda_g$ to $10^4$ (there also exist very small steady-state tracking errors in these two cases). This is expected: the increase of $\lambda_g$ indicates that the optimal control inputs are robust with respect to a larger disturbance set according to Theorem~\ref{thm1}. This leads to conservativeness at the expense of performance if the disturbance set is unnecessarily large. Hence, the tuning of $\lambda_g$ can also be considered as a convenient way to change the response speed of the system.
Conversely, when choosing $\lambda_g$ to be too small, e.g., $\lambda_g = 10^{-5}$, the active power cannot effectively track the reference value. This is because the data in the Hankel matrices contains the true nonlinear behaviors of the system, and leads to prediction errors and inferior performance without proper robustification.

Fig.~\ref{Fig_sim}~(b) shows the active power responses of the converter when considering measurement noise on the output signals. It can be seen that by choosing $\lambda_g = 10$, the converter has effectively perfect tracking performance to the change of power reference. With $\lambda_g = 10^{-5}$, the tracking error is larger than (a) because the disturbance set is not large enough to cover the measurement noise. Fig.~\ref{Fig_sim}~(a) and (b) also plot the A-phase voltage and current of the system with $\lambda_g = 10$. It shows that the sinusoidal waveforms are well controlled by the DeePC algorithm in all the cases.

\subsection{Experiment Results}

In this section, we provide experiment results of the converter system given in Fig.\ref{Fig_control_scheme} to show the performance of the DeePC algorithm in realistic settings. The experiments are based on an Imperix PEB platform (which contains IGBT bridges, passive elements, analog sensors, etc.), and the DeePC algorithm is implemented in an ARM-based processor (B-BOX RCP) which provides control signals to the IGBT bridges \citep{imperix}. The parameters of the experiments are the same as those in the simulation setup. We ignore the input/output constraints and use the closed-form solution of DeePC given in \eqref{eq:closed_form} to enable fast calculation and convenient implementation.

\begin{figure}
\begin{center}
\includegraphics[width=8.0cm]{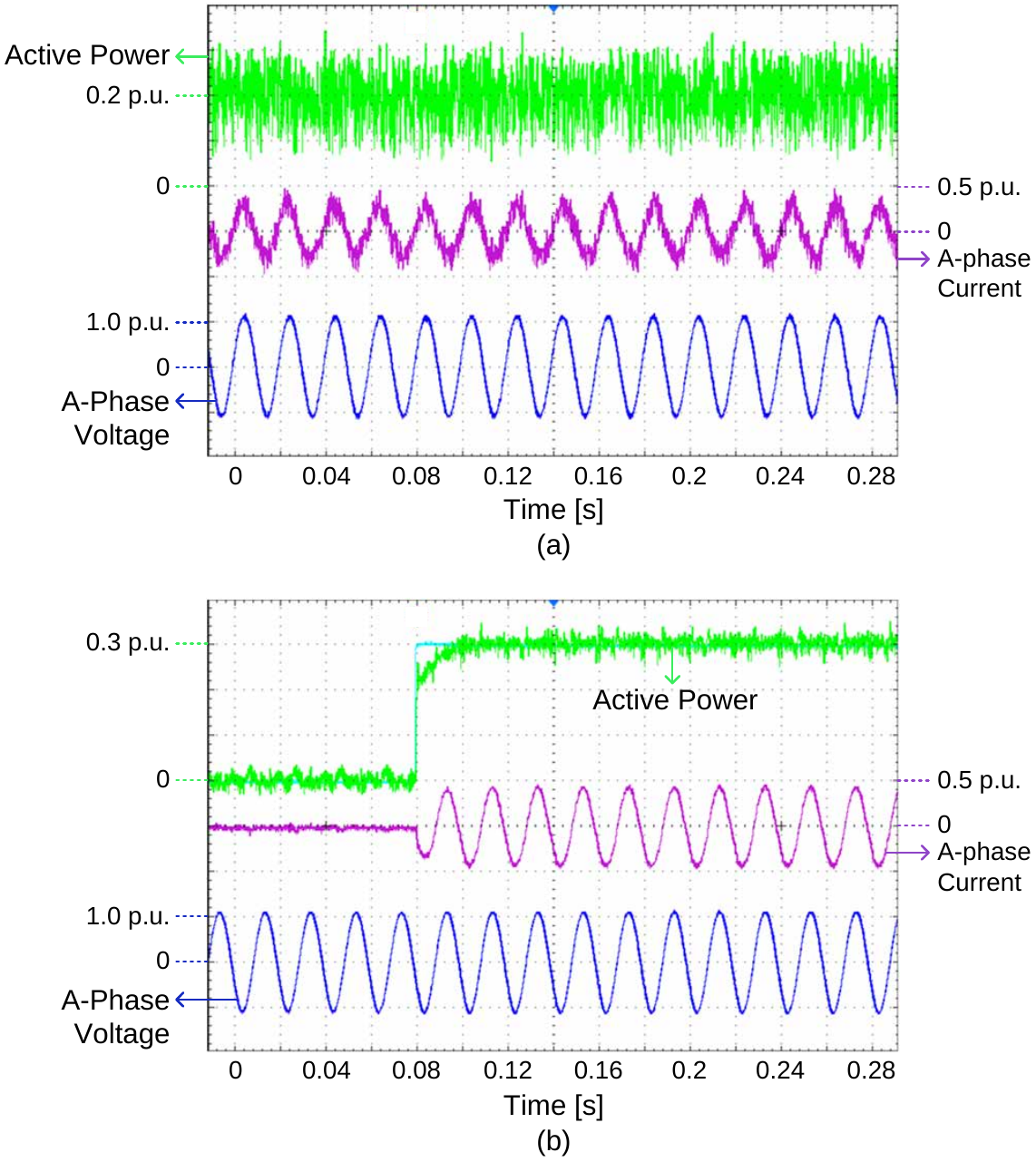}
\vspace{-2mm}
\caption{Experiment results. (a) Responses during the data-collection period (with white-noise signals injected into the system). (b) Responses when the DeePC algorithm is activated, and the active power reference steps from 0 to $0.3{\rm p.u.}$ at $0.08{\rm s}$.}
\vspace{2mm}
\label{Fig_experiment}
\end{center}
\end{figure}

The experiment results are shown in Fig.~\ref{Fig_experiment}. In Fig.~\ref{Fig_experiment}~(a), white-noise signals are injected into the converter such that the input trajectory is persistently exciting. It can be seen that the fluctuations of the active power and current signals are acceptable during this data-collection period.
Fig.~\ref{Fig_experiment}~(b) displays the time-domain responses of the converter when the DeePC algorithm is activated. The active power reference steps from 0 to $0.3{\rm p.u.}$ at $0.08{\rm s}$. It can be seen that the active power perfectly tracks the change of the reference signal, with the rise time being $0.02{\rm s}$. The voltage and current signals are pure sinusoidal signals in steady state, without any power quality issues.



\section{Conclusion}

This paper focused on understanding the effects of the quadratic regularization in DeePC. We explicitly show that the regularized DeePC is equivalent to a min-max formulation which minimizes the worst-case cost for a bounded disturbance set affecting the data Hankel matrices and the initial trajectory. Our analysis highlights the importance of the quadratic regularization in DeePC for robustness. In the absence of input and output constraints, we derived a closed-form solution of the DeePC algorithm with quadratic regularization for fast calculation. Based on the above features, we proposed a framework for grid-connected converters to achieve data-driven synchronization and power regulation, and we tested its performance with high-fidelity simulations and experiments. Future work will investigate the robustness of other forms of regularization and consider structured disturbance sets.

\begin{ack}
The authors would like to thank Catalin Arghir for his help in the experiments.
This research was supported by the ERC under project OCAL (contract number 787845) and ETH Zurich Funds.
\end{ack}

\bibliographystyle{ifacconf-harvard}
\bibliography{ifacconf}


\end{document}